\documentclass[
    ,final            
  ]
  {aipproc}

\layoutstyle{8x11single}

\usepackage{graphicx}
\usepackage{amssymb}
\usepackage{amsmath}

\makeatletter
\def\simleq{\mathrel{\mathpalette\gl@align<}}
\def\simgeq{\mathrel{\mathpalette\gl@align>}}
\def\gl@align#1#2{\lower.6ex\vbox{\baselineskip\z@skip\lineskip\z@
     \ialign{$\m@th#1\hfill##\hfil$\crcr#2\crcr\sim\crcr}}}
\makeatother

\newcommand{\Pu}{p_{\uparrow}}

\newcommand{\Nu}{n_{\uparrow}}
\newcommand{\Nd}{n_{\downarrow}}

\begin{document}

\title{The Lattice QCD Study of the Three-Nucleon Force}



\author{Takumi Doi (for HAL QCD Collaboration)}{
  address={
Center for Nuclear Study, The University of Tokyo,
Tokyo 113-0033, Japan}
}



\begin{abstract}
We investigate three-nucleon forces (3NF)
from lattice QCD simulations,
utilizing the Nambu-Bethe-Salpeter (NBS) wave function
to determine two-nucleon forces (2NF) and 3NF on the same footing.
Quantum numbers of the three-nucleon (3N) system are chosen to be
$(I, J^P)=(1/2,1/2^+)$ (the triton channel).
We consider the simplest geometrical configuration where
3N are aligned linearly with an equal spacing,
to reduce the enormous computational cost. 
Lattice QCD simulations are performed using 
$N_f=2$ dynamical clover fermion configurations
at the lattice spacing of $a = 0.156$ fm on a $16^3\times 32$ lattice
with a large quark mass corresponding to $m_\pi= 1.13$ GeV.
We find repulsive 3NF at short distance.
\end{abstract}

\maketitle


\section{Introduction}
\label{sec:intro}

One of the hottest topic in nuclear physics and astrophysics these days 
is the understanding of the properties of 3NF.
Actually, there are various  phenomena
where 3NF
may play an important role, e.g.,
the binding energies of light nuclei~\cite{Pieper:2007ax},
the properties of neutron-rich nuclei and the supernova nucleosynthesis~\cite{Otsuka:2009cs} 
and
the nuclear equation of state (EoS)
at high density  relevant to the physics of neutron stars~\cite{Akmal:1998cf, Nishizaki:2002ih}.

Despite of its phenomenological importance,
microscopic understanding  of 3NF is  still limited.
Pioneered by Fujita and Miyazawa~\cite{Fujita:1957zz},
3NF have been commonly studied from two-pion exchange (2$\pi$E) models
with the $\Delta$-excitation.
However, since 3NF is originated by the fact that
a nucleon is not a fundamental particle,
it is most desirable to determine 3NF
from the fundamental degrees of freedom (DoF), i.e., quarks and gluons,
on the basis of quantum chromodynamics (QCD).
In this proceeding,
we report the calculation of 3NF from first-principle lattice QCD.

As for the 2NF from lattice QCD,
 an approach based on  the 
 NBS wave function 
 has been  proposed~\cite{Ishii:2006ec, Aoki:2009ji}. 
 Resultant (parity-even) 2NF in this approach
are found to have attractive wells at long and medium
distances  and central repulsive cores at short distance.
The method has been  extended to the
hyperon-nucleon (YN) and hyperon-hyperon (YY) interactions%
~\cite{Nemura:2008sp, Inoue:2010hs, Sasaki:2010bi,Inoue:2010es,Aoki:2011gt}.
In this report, we extend the method to the 3N system,
and perform the lattice QCD simulations of 3NF 
in the triton channel,
$(I, J^P)=(1/2,1/2^+)$~\cite{Doi:2010yh, Doi:baryons2010, Doi:2011gq}.
For details of this study, refer to Ref.~\cite{Doi:2011gq}.

\section{Formalism}
\label{sec:formulation}

Since the detailed formulation for the 2NF is 
given in Ref.~\cite{Aoki:2009ji},
we discuss the extension to the 3N system.
We consider the NBS wave function $\psi_{3N}(\vec{r},\vec{\rho})$ 
extracted from the six-point correlator as
\begin{eqnarray}
\label{eq:6pt_3N}
G_{3N} (\vec{r},\vec{\rho},t-t_0) 
&\equiv& 
\frac{1}{L^3}
\sum_{\vec{R}}
\langle 0 |
          (N(\vec{x}_1) N(\vec{x}_2) N (\vec{x}_3))(t) \
\overline{(N'       N'        N')}(t_0)
| 0 \rangle 
\xrightarrow[t \gg t_0]{} A_{3N} \psi_{3N} (\vec{r},\vec{\rho}) e^{-E_{3N}(t-t_0)} ,
\\
\label{eq:NBS_3N}
\psi_{3N}(\vec{r},\vec{\rho}) &\equiv& 
\langle 0 | N(\vec{x}_1) N(\vec{x}_2) N(\vec{x}_3) | E_{3N}\rangle ,
\qquad
A_{3N} \equiv \langle E_{3N} | \overline{(N' N' N')} | 0 \rangle , 
\end{eqnarray}
where
$E_{3N}$ and $|E_{3N}\rangle$ denote
the energy and the state vector of the 3N ground state, respectively, 
$N$ ($N'$) the nucleon operator in the sink (source),
and
$\vec{R} \equiv ( \vec{x}_1 + \vec{x}_2 + \vec{x}_3 )/3$,
$\vec{r} \equiv \vec{x}_1 - \vec{x}_2$, 
$\vec{\rho} \equiv \vec{x}_3 - (\vec{x}_1 + \vec{x}_2)/2$
the Jacobi coordinates.

With the derivative expansion of the potentials~\cite{Murano:2011nz},
the NBS wave function can be converted to the potentials
through the following 
Schr\"odinger equation,
\begin{eqnarray}
%
\biggl[ 
- \frac{1}{2\mu_r} \nabla^2_{r} - \frac{1}{2\mu_\rho} \nabla^2_{\rho} 
+ \sum_{i<j} V_{2N} (\vec{r}_{ij})
+ V_{3NF} (\vec{r}, \vec{\rho})
\biggr] \psi_{3N}(\vec{r}, \vec{\rho})
= E_{3N} \psi_{3N}(\vec{r}, \vec{\rho}) , 
\label{eq:Sch_3N}
\end{eqnarray}
where
$V_{2N}(\vec{r}_{ij})$ with $\vec{r}_{ij} \equiv \vec{x}_i - \vec{x}_j$
denotes the 2NF between $(i,j)$-pair,
$V_{3NF}(\vec{r},\vec{\rho})$ the 3NF,
$\mu_r = m_N/2$, $\mu_\rho = 2m_N/3$ the reduced masses.
If we calculate 
$\psi_{3N}(\vec{r}, \vec{\rho})$ for all $\vec{r}$ and $\vec{\rho}$,
and if all $V_{2N}(\vec{r}_{ij})$ are obtained
by (separate) lattice calculations for genuine 2N systems,
we can extract $V_{3NF}(\vec{r},\vec{\rho})$ through Eq.~(\ref{eq:Sch_3N}).

In practice, however, the computational cost is enormous,
because of enlarged DoF by the 3N (i.e., 9 quarks) and 
factorial number of Wick contractions.
In order to reduce the cost, 
we develop several techniques, 
e.g., taking advantage of symmetries, and
employing the non-relativistic limit for the nucleon operator in the source.
We further restrict the geometry of the 3N.
More specifically, we consider the ``linear setup''with $\vec{\rho}=\vec{0}$,
with which 3N are aligned linearly with equal spacings of 
$r_2 \equiv |\vec{r}|/2$.
%
In this setup,
the third nucleon is attached
to $(1,2)$-nucleon pair with only S-wave.
Considering the total 3N quantum numbers of 
$(I, J^P)=(1/2,1/2^+)$,
the triton channel, 
the wave function can be completely spanned by
only three bases, which can be labeled
by the quantum numbers of $(1,2)$-pair as
$^1S_0$, $^3S_1$, $^3D_1$.
Therefore, the Schr\"odinger equation
leads to 
the $3\times 3$ coupled channel equations
with the bases of 
$\psi_{^1S_0}$, $\psi_{^3S_1}$, $\psi_{^3D_1}$.
The reduction of the dimension of bases 
is expected to improve the S/N as well.

We then consider the identification of genuine 3NF.
It is a nontrivial work:
Although both of parity-even and parity-odd 2NF
are required to subtract 2NF part in Eq.~(\ref{eq:Sch_3N}),
parity-odd 2NF have not been obtained yet in lattice QCD.
In order to resolve this issue,
we consider the following channel,
\begin{eqnarray}
\psi_S \equiv
\frac{1}{\sqrt{6}}
\Big[
-   \Pu \Nu \Nd + \Pu \Nd \Nu               
                - \Nu \Nd \Pu + \Nd \Nu \Pu 
+   \Nu \Pu \Nd               - \Nd \Pu \Nu
\Big]  ,
\label{eq:psi_S}
\end{eqnarray}
which is anti-symmetric
in spin/isospin spaces 
for any 2N-pair.
Combined with the Pauli-principle,
it is automatically guaranteed that
any 2N-pair couples with even parity only.
Therefore, we can extract 3NF unambiguously 
using only parity-even 2NF.
Note that no assumption on the choice of 3D-configuration of $\vec{r}$, $\vec{\rho}$
is imposed in this argument,
and we can take advantage of this feature
for future 3NF calculations with various 3D-configuration setup.

\section{Lattice QCD setup and results}
\label{sec:results}

We employ
$N_f=2$ dynamical 
configurations
with mean field improved clover fermion 
and 
RG-improved
gauge action
generated by CP-PACS Collaboration~\cite{Ali Khan:2001tx}.
We use
598 configurations at
$\beta=1.95$ and
the lattice spacing of
$a^{-1} = 1.269(14)$ GeV,
and 
the lattice size of $V = L^3 \times T = 16^3\times 32$
corresponds to
(2.5 fm)$^3$ box in physical spacial size.
For $u$, $d$ quark masses, 
we take the hopping parameter at the unitary point
as
$\kappa_{ud} = 0.13750$,
which corresponds to
$m_\pi = 1.13$ GeV, 
$m_N = 2.15$ GeV and
$m_\Delta = 2.31$ GeV.
We use the wall quark source with Coulomb gauge fixing.
In order to enhance the statistics,
we perform the measurement at 32 source time slices for each configuration,
and the forward and backward propagations are averaged.
The results from both of 
total angular momentum $J_z=\pm 1/2$ 
are averaged as well.
We perform the simulation
at eleven physical points of the distance $r_2$ with the linear setup.

\begin{figure}[t]
\begin{minipage}{0.5\textwidth}
\begin{center}
%
\includegraphics[width=0.85\textwidth]{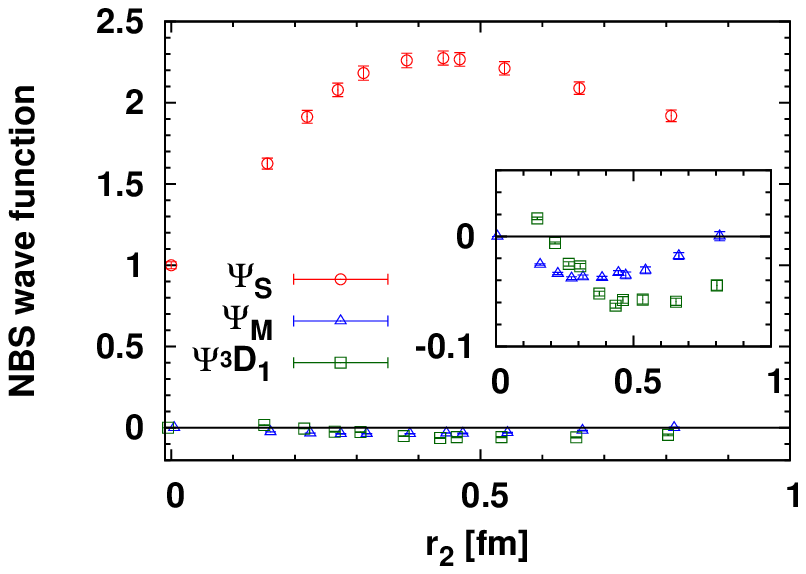}
%
%
\end{center}
\end{minipage}
\hfill
\begin{minipage}{0.5\textwidth}
\begin{center}
%
\includegraphics[width=0.85\textwidth]{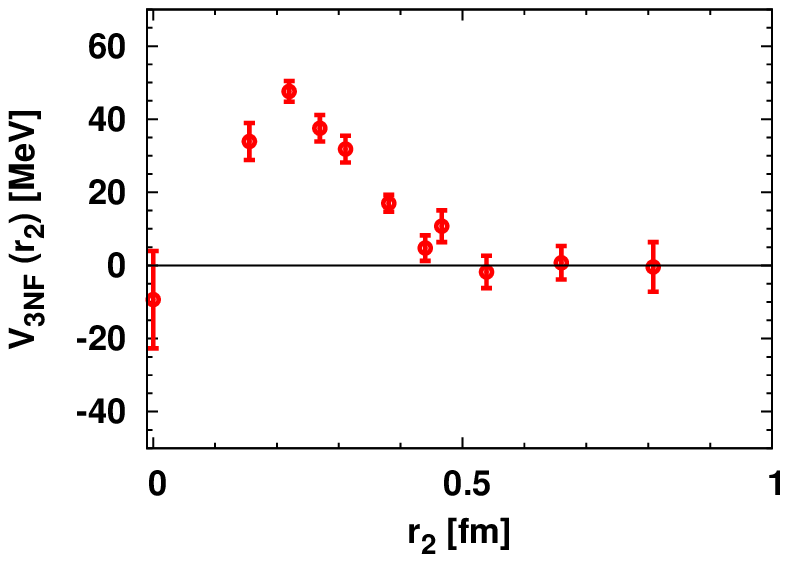}
\caption{
\label{fig:3N}
(color online).
(Left) 
3N wave functions 
at $(t-t_0)/a=8$.
Circle (red), triangle (blue), square (green) points denote
$\psi_S$, $\psi_M$, $\psi_{\,^3\!D_1}$, respectively.
(Right) 
The effective scalar-isoscalar 3NF 
in the triton channel with the linear setup.
%
\vspace*{-5mm}
}
\end{center}
\end{minipage}
\end{figure}

In Fig.~\ref{fig:3N} (left),
we plot
the radial part of each wave function of
$\psi_S = ( - \psi_{^1S_0} + \psi_{^3S_1} )/\sqrt{2}$,
$\psi_M \equiv ( \psi_{^1S_0} + \psi_{^3S_1} )/\sqrt{2}$
and
$\psi_{^3D_1}$ 
obtained at $(t-t_0)/a = 8$.
Here, we normalize the wave functions
by the center value of $\psi_S(r_2=0)$.
What is noteworthy is that
the wave functions are obtained with good precision,
which is quite nontrivial for the 3N system.
We observe that 
$\psi_S$ overwhelms the wave function,
indicating that higher partial waves 
are strongly suppressed,
and thus the effect of the next leading order in the derivative expansion,
spin-orbit forces,
is suppressed in this lattice setup.

We determine 3NF
by subtracting 2NF from total potentials in the 3N system.
Since we have only one channel (Eq.~(\ref{eq:psi_S})) 
which is free from parity-odd 2NF,
we can determine one type of 3NF.
In this report,
3NF are
effectively represented in 
a scalar-isoscalar functional form,
which is 
often employed for the 
short-range 3NF in phenomenology.

In Fig.~\ref{fig:3N} (right), we plot the results
for the effective scalar-isoscalar 3NF at $(t-t_0)/a = 8$.
Here, 
we include $r_2$-independent shift by energies,
$\delta_E \simeq 5$~MeV,
which is determined by 
long-range behavior of potentials (2NF and effective 2NF in the 3N system)~\cite{Doi:2011gq}.
While $\delta_E$ suffers from $\simleq 10$ MeV systematic error,
it does not affect the following discussions much, since $\delta_E$ merely serves as an overall offset.
In order to check the dependence on the sink time slice,
we calculate 3NF at $(t-t_0)/a =$ 9 as well,
and confirm that the results are consistent with each other~\cite{Doi:2011gq}.

Fig.~\ref{fig:3N} (right) shows that
3NF are small at the long distance region of $r_2$.
This is in accordance with the suppression
of 2$\pi$E-3NF by the heavy pion.
At the short distance region, on the other hand,
an indication of repulsive 3NF is observed.
Note that a repulsive short-range 3NF 
is phenomenologically required 
to explain the properties of high density matter.
Since multi-meson exchanges are strongly suppressed 
by the large quark mass, 
the origin of this short-range 3NF may be attributed to the 
quark and gluon dynamics directly.
In fact, we recall that the short-range repulsive (or attractive) cores
in the generalized two-baryon potentials 
are 
calculated in lattice QCD in the flavor SU(3) limit, 
and the results are found to be well explained 
from the viewpoint of the Pauli exclusion principle in the quark level~\cite{Inoue:2010hs}.
In this context, 
it is intuitive to expect that the 3N system is subject to extra Pauli repulsion effect,
which could be an origin of the observed short-range repulsive 3NF.
Further investigation along this line is certainly an interesting subject in future.

As regards the systematic error,
one may worry about the discretization error,
since the nontrivial results are obtained at short distance.
In particular, 
the kinetic terms 
could suffer from
a substantial effect, since they are calculated by the finite 
difference Laplacian operator as
$
\nabla^2 f(x) = 
\nabla^2_{\rm std} f(x) \equiv
\frac{1}{a^2}
\sum_{i}
\left[ f(x+a_i) + f(x-a_i) - 2 f(x) \right] 
$.
In order to estimate this artifact,
we also analyze using the improved Laplacian operator
for both of 2N and 3N, 
$
\nabla^2_{\rm imp} f(x)
\equiv
\frac{1}{12a^2}
\sum_{i}
\left[
-    ( f(x+2a_i) + f(x-2a_i) )
+ 16 ( f(x+a_i)  + f(x-a_i)  )
-30   f(x))
\right] 
$.
We observe that 
the results are consistent with each other,
and the discretization artifact of 3NF in Laplacian operator is small~\cite{Doi:2011gq}.
Of course, this study probes only a part of discretization errors,
and explicit simulations with a finer lattice are on-going.

Since the lattice simulations are carried out 
only at single large quark mass,
quark mass dependence of 3NF 
is certainly 
an important issue. 
In the case of 2NF,
short-range cores have the enhanced strength
and broaden range by decreasing the mass%
~\cite{Aoki:2009ji}.
We, therefore, would expect a significant quark mass dependence
exist in short-range 3NF as well.
Quantitative investigation through
lattice simulations with lighter quark masses
are currently underway.


We thank authors and maintainers of CPS++\cite{CPS}.
We also thank  
CP-PACS Collaboration
and ILDG/JLDG~\cite{conf:ildg/jldg} for providing gauge configurations.
The numerical simulations have been performed
on Blue Gene/L at KEK,
T2K at University of Tsukuba and SR16000 at YITP in Kyoto University.
This research is supported in part by MEXT Grant-in-Aid (20340047, 22540268),
Scientific Research on Innovative Areas (20105001, 20105003, 21105515),
Specially Promoted Research (13002001), JSPS 21$\cdot$5985 and HPCI PROGRAM,
the Large Scale Simulation Program of KEK (09-23, 09/10-24) and the collaborative
interdisciplinary program at T2K-Tsukuba (09a-11, 10a-19).


\vspace*{-3mm}
\begin{thebibliography}{99}

\bibitem{Pieper:2007ax}
  S.~C.~Pieper,
  Riv.\ Nuovo Cim.\  {\bf 31}, 709 (2008)
  [arXiv:0711.1500 [nucl-th]].

\bibitem{Otsuka:2009cs}
  T.~Otsuka, T.~Suzuki, J.~D.~Holt, A.~Schwenk and Y.~Akaishi,
  Phys.\ Rev.\ Lett.\  {\bf 105}, 032501 (2010)
  [arXiv:0908.2607 [nucl-th]].

\bibitem{Akmal:1998cf}
  A.~Akmal, V.~R.~Pandharipande and D.~G.~Ravenhall,
  Phys.\ Rev.\  {\bf C58}, 1804 (1998)
  [nucl-th/9804027].

\bibitem{Nishizaki:2002ih}
  S.~Nishizaki, T.~Takatsuka and Y.~Yamamoto,
  Prog.\ Theor.\ Phys.\  {\bf 108}, 703 (2002).\\
  T.~Takatsuka, S.~Nishizaki and R.~Tamagaki,
  Prog.\ Theor.\ Phys.\ Suppl.\  {\bf 174}, 80 (2008).

\bibitem{Fujita:1957zz}
  J.~Fujita and H.~Miyazawa,
  Prog.\ Theor.\ Phys.\  {\bf 17}, 360 (1957).

\bibitem{Ishii:2006ec}
  N.~Ishii, S.~Aoki and T.~Hatsuda,
  Phys.\ Rev.\ Lett.\  {\bf 99}, 022001 (2007)
  [nucl-th/0611096].

\bibitem{Aoki:2009ji}
  S.~Aoki, T.~Hatsuda and N.~Ishii,
  Prog.\ Theor.\ Phys.\  {\bf 123}, 89 (2010)
  [arXiv:0909.5585 [hep-lat]].

\bibitem{Nemura:2008sp}
  H.~Nemura, N.~Ishii, S.~Aoki and T.~Hatsuda,
  Phys.\ Lett.\  {\bf B673}, 136 (2009)
  [arXiv:0806.1094 [nucl-th]].

\bibitem{Inoue:2010hs}
  T.~Inoue {\it et al.} [HAL QCD Collab.],
  Prog.\ Theor.\ Phys.\  {\bf 124}, 591 (2010)
  [arXiv:1007.3559 [hep-lat]].

\bibitem{Sasaki:2010bi}
  K.~Sasaki [HAL QCD Collab.],
  PoS {\bf LATTICE2010}, 157 (2010)
  [arXiv:1012.5685 [hep-lat]].

\bibitem{Inoue:2010es}
  T.~Inoue {\it et al.} [HAL QCD Collab.],
  Phys.\ Rev.\ Lett.\  {\bf 106}, 162002 (2011)
  [arXiv:1012.5928 [hep-lat]].

\bibitem{Aoki:2011gt}
  S.~Aoki {\it et al.}  [HAL QCD Collaboration],
  Proceedings of the Japan Academy, Ser. B, in press,
  [arXiv:1106.2281 [hep-lat]].

\bibitem{Doi:2010yh}
  T.~Doi [HAL QCD Collab.],
  PoS {\bf LATTICE2010}, 136 (2010)
  [arXiv:1011.0657 [hep-lat]].

\bibitem{Doi:baryons2010}
  T.~Doi [HAL QCD Collab.], 
  Proc. of the Int. Conf. on the structure of baryons (BARYONS'10),
  [arXiv:1105.6247 [hep-lat]].

\bibitem{Doi:2011gq}
  T.~Doi {\it et al.} [HAL QCD Collab.],
  arXiv:1106.2276 [hep-lat].

\bibitem{Murano:2011nz}
  K.~Murano, N.~Ishii, S.~Aoki and T.~Hatsuda,
  Prog.\ Theor.\ Phys.\  {\bf 125}, 1225 (2011)
  [arXiv:1103.0619 [hep-lat]].



\bibitem{Ali Khan:2001tx}
  A.~Ali Khan {\it et al.}  [CP-PACS Collab.],
  Phys.\ Rev.\  D {\bf 65}, 054505 (2002)
  [E:\  D {\bf 67}, 059901 (2003)].


\bibitem{CPS}
Columbia Physics System (CPS), \url{http://qcdoc.phys.columbia.edu/cps.html}

\bibitem{conf:ildg/jldg}
  \url{http://www.lqcd.org/ildg},
  \url{http://www.jldg.org}


\end{thebibliography}
\end{document}